# Ground-based near-infrared observations of water vapour in the Venus troposphere.

**S.Chamberlain[1], J.A.Bailey,[2] D.Crisp,[3] and V.S.Meadows[4]**

[1]Centre of Astronomy and Astrophysics, University of Lisbon, Portugal.
[2]School of Physics, University of New South Wales, Sydney, Australia
[3]Jet Propulsion Laboratory, California Institute of Technology, Pasadena, California, USA.
[4]Department of Astronomy, University of Washington, Seattle, Washington, USA.
*Email: DrSarahChamberlain@gmail.com*

**Abstract:**
We present a study of water vapour in the Venus troposphere obtained by modelling specific water vapour absorption bands within the 1.18 μm window. We compare the results with the normal technique of obtaining the abundance by matching the peak of the 1.18 μm window. Ground-based infrared imaging spectroscopy of the night side of Venus was obtained with the Anglo-Australian Telescope and IRIS2 instrument with a spectral resolving power of R~2400. The spectra have been fitted with modelled spectra simulated using the radiative transfer model VSTAR. We find a best fit abundance of 31 ppmv (-6 + 9 ppmv), which is in agreement with recent results by Bézard et al. 2011 using VEX/SPICAV (R~1700) and contrary to prior results by Bézard et al. 2009 of 44 ppmv (+/-9 ppmv) using VEX/VIRTIS-M (R~200) data analyses. Comparison studies are made between water vapour abundances determined from the peak of the 1.18 μm window and abundances determined from different water vapour absorption features within the near infrared window. We find that water vapour abundances determined over the peak of the 1. 18 μm window results in plots with less scatter than those of the individual water vapour features and that analyses conducted over some individual water vapour features are more sensitive to variation in water vapour than those over the peak of the 1. 18 μm window. No evidence for horizontal spatial variations across the night side of the disk are found within the limits of our data with the exception of a possible small decrease in water vapour from the equator to the north pole. We present spectral ratios that show water vapour absorption from within the lowest 4 km of the Venus atmosphere only, and discuss the possible existence of a decreasing water vapour concentration towards the surface.

## 1.1 Introduction:

Water vapour is a strong greenhouse gas, the determination of its profile and distribution in the lower atmosphere is important for understanding its role in maintaining the high surface temperatures. Water vapour is an important chemical reactant in the lower atmosphere as it is the major reservoir of hydrogen, which is hypothesized to buffer or regulate the atmospheric abundances of HCL and HF (Fegley, 2003). Water vapour is important to the formation of $H_2SO_4$ clouds (Krasnopolsky and Pollack, 1994) and also pertinent to surface – atmosphere interactions. The main sources of water vapour replenishment are believed to be volcanic outgassing and cometary impacts (Fegley, 2003). Water vapour is lost on long timescales through oxidation reactions with iron minerals at the Venus surface (Fegley, 2003) as well as through photo-disassociation after water vapour is transported into the upper atmosphere. Therefore studies of the abundance distributions and profiles of water vapour in the lower troposphere help to constrain the chemistry and evolution of the Venus near surface environment.

The Venus lower atmosphere is difficult to study by remote observations or *in situ* probes, due to inhospitable surface conditions and a global layer of dense cloud that extends from approximately 70 km altitude down to 45 km - 50 km altitude with a lower thin haze layer that extends from the base of the clouds to approximately 30 km and an upper thin haze layer that extends from the cloud top to 90 km altitude (Esposito et al. 1997). Early *in situ* instruments sent back a confusing mix of results with respect to the Venus water vapour abundance profile below the cloud. These early results are summarised by Meadows and Crisp (1996). The discovery of near-infrared windows in the night side spectrum of Venus (Allen and Crawford, 1984) opened up a new approach to the study of the deep atmosphere. These windows occur between strong $CO_2$ and $H_2O$ absorption bands at wavelengths where the sulphuric acid clouds scatter light, but only weakly absorb (Pollack *et al.* 1993).

Near infrared windows are observed on the Venus nightside where the scattered daylight radiation is minimal allowing thermal radiation emission from the deep lower atmosphere to be detected. Near-infrared windows at different wavelengths originate at different altitudes in the Venus atmosphere (Crisp *et al.* 1991a, Pollack *et al.* 1993 and Meadows and Crisp 1996). The near-infrared window at 2.3 µm probes an altitude of 35 km, just below the clouds, 1.74 µm is sensitive to approximately 25 km and the 1.18 µm window is influenced by changes in the temperature and composition at altitudes near 15 km. The above mentioned windows are sensitive to water vapour abundance and can therefore be used to probe the water vapour abundance profile below the cloud deck. Previous remote observations have provided a more consistent set of results than those of *in situ* measurements (*in situ* measurements are summarised in Meadows and Crisp, 1996). The current consensus is that the water vapour abundance in the lower atmosphere of Venus is constant from about 30 km altitude to the surface with an abundance of approximately 30 ppmv. This agrees with predictions made by recent chemical models (Krasnopolsky, 2007), however the error bars are relatively large and allow for a range of values and possible trends in near surface gradient. Results to date are summarised in table 1.

Prior to 2010 no horizontal variations were detected (Drossart et al. 1993, Meadows and Crisp (1996)), with the exception of Bell et al. (1991), who discovered anomalously strong variations in water vapour abundance between (40 – 200 ppm) using the 2.3 µm window corresponding to approximately 35 km, however this strength of variation has not been seen since. Bézard et al. (2009) placed an upper limit on variations in H2O abundance using the 1.18 µm window (15 – 20 km altitude) of +/-1% between 60 S and 25 N and +/-2% for 80 S to 25 N derived from VEX/VIRTIS-M data.

Tsang et al. (2010) have recently published data from the VIRTIS imaging spectrometer that indicates a variation of 22 to 35 ppmv at 30 – 40 km altitude. These variations are anticorrelated with cloud opacity, supporting the mechanism originally proposed by Bell et al. (1991) that the observed variations in H2O are associated with the rainout and subsequent disassociation of H2SO4 producing H2O and SO3. Large variations in H2SO4 have been observed at 35 – 45 km (Kolodner and Steffes 1998) supporting this observation. That these variations have not been observed in the 1.18 µm window previously indicates that by 30 km altitude the Venus troposphere is globally well mixed.

Tsang et al. (2010) have recently published data from the VIRTIS imaging spectrometer that indicates a variation of 22 to 35 ppmv at 30 – 40 km altitude. These variations are anticorrelated with cloud opacity, supporting the mechanism originally proposed by Bell et al. (1991) that the observed variations in H2O are associated with the rainout and subsequent disassociation of H2SO4 producing H2O and SO3. Large variations in H2SO4 have been observed at $35 - 45$ km (Kolodner and Steffes 1998) supporting this observation. That these variations have not been observed in the 1.18 µm window previously indicates that by 30 km altitude the Venus troposphere is globally well mixed.

Table 1: Venusian subcloud water vapour abundances obtained remotely by ground-based and spacecraft observations. Table from Bailey (2009) and later amended.

| Atmospheric Window | Water Vapour, ppmv | | Reference | Instrument |
|---|---|---|---|---|
| 2.3µm (30 – 45km) | 40 | | Bézard *et al.* (1990) | CFHT/FTS |
| | 25 | +25/-13 | Carlson *et al.* (1991) | Galileo NIMS |
| | 40 | +/-20 | Crisp *et al.* (1991a) | AAT/FIGS |
| | 40 | | Bell *et al.* (1991) | IRTF/CGAS - Dry profile |
| | 200 | | Bell *et al.* (1991) | IRTF/CGAS - Wet profile |
| | 30 | +/-6 | Pollack *et al.* (1993) | AAT/FIGS |
| | 30 | +15/-10 | De Bergh *et al.* (1995) | CFHT/FTS |
| | 26 | +/-4 | Marcq *et al.* (2006) | IRTF/SPEX |
| | 31 | +/-2 | Marcq *et al.* (2008) | VEX/VIRTIS-H |
| | 22 – 35 | +/-4 | Tsang et al. (2010) | VEX/VIRTIS |
| 1.74µm (15 – 25km) | 50 | +50/-25 | Carlson *et al.* (1991) | Galileo NIMS |
| | 40 | | Bézard *et al.* (1990) | CFHT/FTS |
| | 30 | +/-7.5 | Pollack *et al.* (1993) | AAT/FIGS |
| | 30 | +/-10 | De Bergh *et al.* (1995) | CFHT/FTS |
| 1.28µm | 40 | +/-20 | Crisp *et al.* (1991a) | AAT/FIGS |
| 1.18µm (5 - 15km) | 30 | +/- 15 | Drossart *et al.* (1993) | Galileo NIMS |
| | 30 | +/-15 | De Bergh *et al.* (1995) | CFHT/FTS |
| | 45 | +/-10 | Meadows and Crisp (1996) | AAT/IRIS |
| | 44 | +/-9 | Bézard *et al.* (2009) | VEX/VIRTIS-M |
| | 30 | +/10-5 | Bézard *et al.* (2011) | VEX/SPICAV |

Previous studies to determine the abundance of water vapour using the 1.18 µm window were made by modelling the shape of this window or relative height of this window with respect to the 1.01 µm or 1.27 µm window (Drossart *et al.* 1993, Meadows and Crisp 1996, Bézard *et al.* 2009, 2011). Prior to VEX/SPICAV (resolving power R (=λ/∆λ) ∼ 1700) most observations were obtained at spectral resolutions too low to attempt modelling of individual water vapour absorption features. Whilst the water vapour features are observed in VEX/SPICAV observations, water vapour abundances were determined based on the overall shape of the window which is complicated by the contribution of $CO_2$ absorption. Here we use spectra with a resolving powers of R ~ 2400 and present the first analysis of water vapour in the troposphere of Venus obtained by matching individual water vapour absorption features within the 1.18 μm window. Although the individual water vapour absorption lines are not fully resolved in our observed spectra, all the lines within each feature are due to $H_2O$, and thus they are a good choice for determining $H_2O$ abundance.

## 1.2 Observations and Data Reduction:

Near-infrared observations of the Venus night side were obtained using the InfraRed Imaging Spectrometer 2 (IRIS2) on the 3.9 m Anglo-Australian Telescope (AAT) at Siding Spring Observatory, Australia. Data presented here were obtained on July 12[th] (scan 222) and 14[th] 2004 (Scan 523) (Table 2). The 2004 July observations were chosen out of a number of available datasets because they were obtained in relatively cold conditions (morning twilight in southern winter) and thus have a relatively low column of telluric water vapour absorption, which can be a significant problem at the low altitude Siding Spring site.

Table 2: Lists the observing parameters for the data obtained for this study

|  | Scan 222 | Scan 523 |
|---|---|---|
| Date of Observation | 12 July 2004 | 14 July 2004 |
| Time of Observation start (UT) | 20:34:53 | 20:37:00 |
| Time of Observation end (UT) | 20:49:52 | 20:51:57 |
| Airmass at start of scan | 2.262 | 2.196 |
| Airmass at finish of scan | 2.097 | 2.054 |
| Angular diameter (arcsec) | 38.34 | 37.15 |
| Illuminated Fraction (%) | 25.25 | 26.98 |
| Sub-Earth Longitude | 20.42 | 24.29 |
| Sub-Earth latitude | 3.78 | 3.81 |
| Spatial Resolution | 69.6km / pixel | 71.2km / pixel |
|  |  |  |
| Slit width | 1 arcsec | |
| Slit length | 7.5 arcmin | |
| Wavelength start | 1.10 μm | |
| Wavelength finish | 1.33 μm | |
| Spectral resolution | 2400 | |
| Exposure time | 5 sec | |
| Step size | 0.446 arcsec | |
| Average Seeing for Site | 1.2 arcsec | |

The observing methods used for this work have been previously described by Bailey *et al.* (2008a, 2008b). Spectral imaging observations were obtained by orienting the spectrometer slit perpendicular to the solar vector and stepping the slit across the Venus disk obtaining a spectrum at each slit position. Each scan began off the Venus dark side and was then stepped incrementally across the Venus disk until the sunlit cusps were visible. The scan was then reversed and stepped back across the disk to the start position. Each scan (forward and reverse) took approximately 15 min to complete.

Data from each scan were constructed into a data cube with 2 spatial axes and one spectral axis, the slit length (y axis), slit scan direction (x axis) and spectral data (z axis). Flat fielding and sky subtraction were both applied to the data during the cube building process. Sky subtraction is particularly important for Venus since observations are obtained just before sunrise when the sky brightness is changing rapidly. Sky frames were obtained at the beginning and end of each scan where the slit is positioned off the darkened side of the Venus disk. These first and last frames were averaged and used as the sky subtraction frame, however a long slit was used for the Venus scans such that spectra obtained at the top and bottom of the slit were far enough away from Venus to be used as a scale for sky brightness. The sky frame was therefore scaled to match the sky brightness at each slit step of the scan and subtracted from the data individually to adjust for the rapid variations in brightness across each scan. Due to this procedure, the dark and bias adjustments were done separately to sky subtraction. Spectral wavelength calibrations were made using a Xenon lamp.

Scattered light from the illuminated crescent causes a brightness gradient across the Venus night side. This effect can be removed by obtaining a pure crescent spectrum from near the saturated crescent immediately off the planet's disk that includes no night side thermal emission (1.22 – 1.25 µm). This crescent spectrum is scaled to a spectral region chosen from the night side spectra where the thermal emission contribution is effectively zero. The scaled crescent spectrum is then subtracted from the night side spectra and this process is repeated for every pixel location.

Water vapour is present in both the Venus atmosphere and the terrestrial atmosphere. Removal of telluric contributions is normally achieved by dividing the data by that of a standard star observed on the same night of observations; however this may introduce several additional errors. Bailey *et al.* (2007) show that systematic errors can be introduced using this process due to unresolved saturated absorption bands by up to 50 % in the case of $CO_2$ in the Martian atmosphere. Further complications are caused by unresolved Doppler shifts of Venusian absorption bands with respect to terrestrial absorption bands and shifts caused by instrumental flexing at large zenith angles. For greater accuracy we therefore approach these problems by including the terrestrial contribution, Doppler shifts and instrumental errors in our modelling rather than dividing our observations by resolution limited data.

Further errors may be introduced using the standard star technique as the standard observations are usually made before or after a Venus scan, and are rarely found at the same zenith angle as that of Venus. Terrestrial water vapour abundances may vary over the course of the night and the abundance of water vapour determined for the terrestrial atmosphere will therefore not accurately represent the terrestrial water vapour abundance at the time of the Venus observations. Whilst the zenith angle effect can be allowed for in modelling, we cannot account for temporal variations in terrestrial water vapour abundance. The solution for this is to use the tops of the Venus clouds that reflect a large percentage of incident sunlight. This sunlight is reflected from pressures < 100 hPa, where the typical water vapour mixing ratios are ~ 1 ppm, so that Venus water contributes little or no absorption to the crescent spectra at these wavelengths. The solar spectrum is well known and this is used in a similar way to that of a standard star. By extracting a scattered crescent spectrum a short distance off the Venus disk we obtain a solar spectrum that has passed through a simultaneous and almost identical terrestrial atmospheric path length to that of the Venus nightside observations. The scattered spectra therefore depict a true abundance of the water vapour in the terrestrial atmosphere at each time step within the scan. The solar spectrum is then passed as an input parameter into the model of the terrestrial atmosphere. Various values of terrestrial water vapour abundances are modelled until the best fit transmittance spectrum is found (Fig 1). The nominal terrestrial water vapour profile is taken from the standard mid-latitude clear sky terrestrial atmospheric profile by the Intercomparison of Radiation Codes in Climate Models (ICRCCM) program. The lower levels are adjusted to match conditions at the Siding Spring site. The Terrestrial water vapour abundance in the lower levels are then varied within the radiative transfer model as a percentage of the nominal water vapour abundance and is therefore referred to as a percentage of the nominal profile rather than specified at each atmospheric level as an abundance in ppm (parts per million).

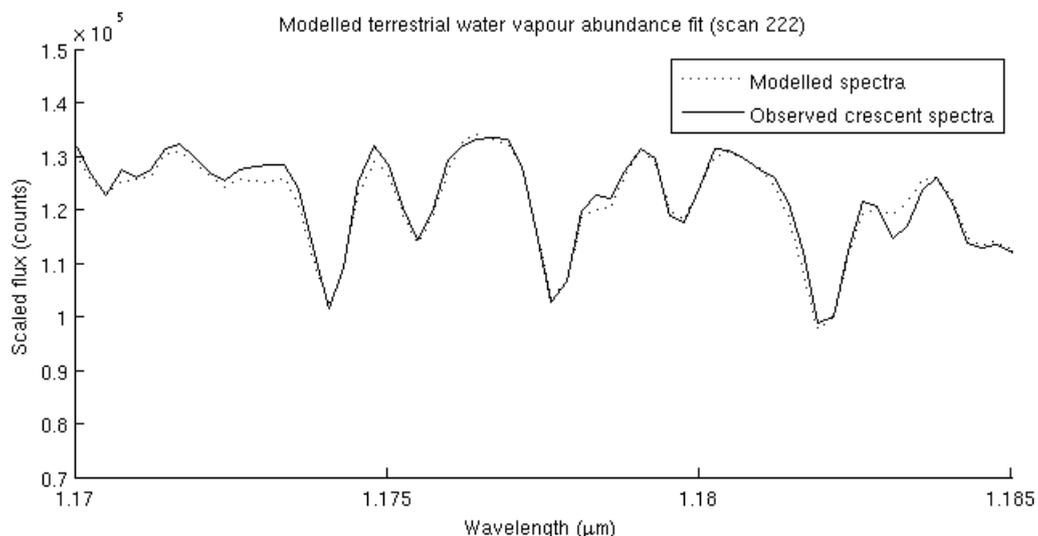

Fig 1: Shows the modelled terrestrial transmission spectrum (dashed line) matched to an observed crescent spectrum (solid line) for scan 222.

Over each 15 minute scan, best fit water vapour abundances were found for each slit position corresponding to Venus nightside spectral observations in order to verify that the terrestrial water vapour profile did not vary over the scan time frame (Fig. 2).

Modelling showed that an error of 1% in the terrestrial water vapour content corresponds to a variation of approximately 0.3 – 0.7 ppmv in the determined Venusian water vapour abundance ( 0.3 ppmv for feature 1 and the peak feature and 0.7 ppmv for feature 2 and feature 3, see fig 3 for feature references). The modelled results also indicate that the best fit Venus water vapour profile is more strongly influenced by water vapour variations in the Venus atmosphere than those in the Terrestrial atmosphere and that features 2 and 3 are more sensitive to the Terrestrial water vapour profile than feature 1 and peak (shown in section 1.3 fig. 4). Figure 2 shows the relative temporal variations in the Terrestrial water vapour profile, with respect to the nominal model, by plotting the best fit modelled terrestrial water vapour percentage match at each spectral slit position (each slit location that corresponds to a Venus disk spectrum position). As pixels close to the edge of the disk are not included as a part of the analysis and the observed variations do not introduce significant errors in the determined Venus water vapour profile, we assume a constant terrestrial water vapour profile over the time it took to complete each scan.

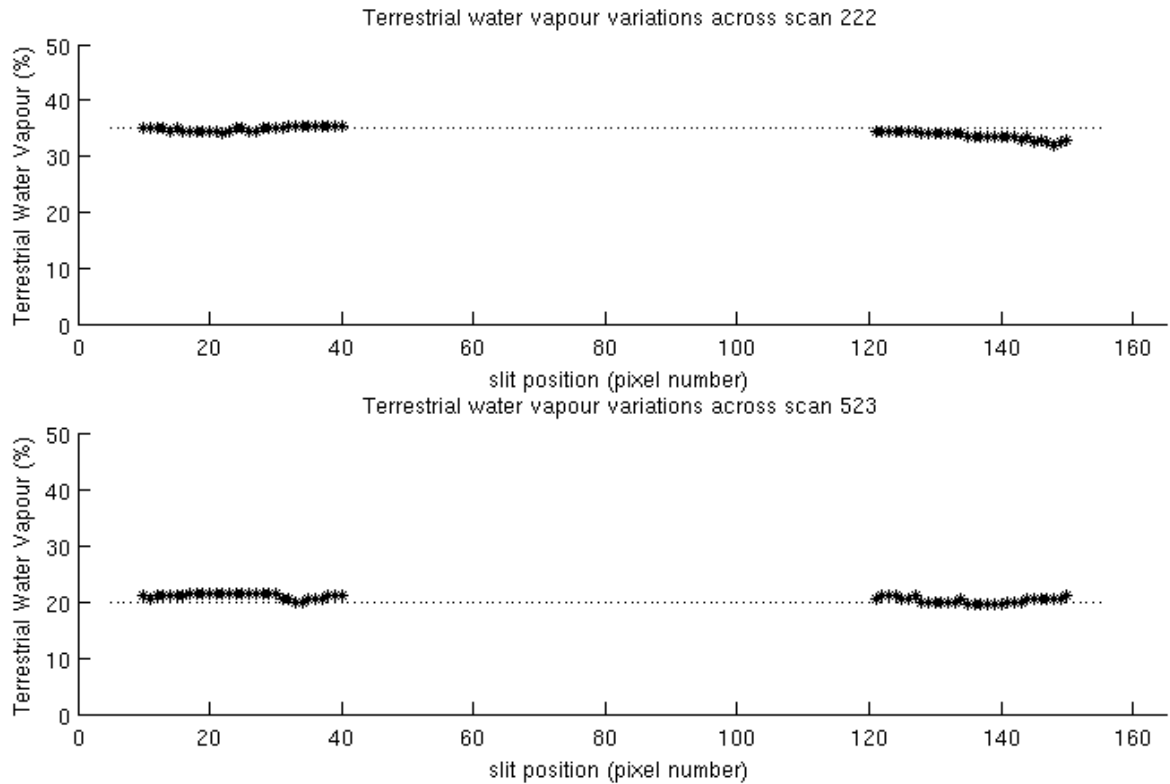

Fig 2: Shows best fit water vapour profile percentage for slit positions that correspond to Venus night side spectra. The water vapour profile percentage is taken relative to the standard mid-latitude clear sky terrestrial atmospheric profile, provided by the Intercomparison of Radiation Codes in Climate Models (ICRCCM) program, with the lower levels are adjusted to match conditions at the Siding Spring site. Variations are shown for both the forward and reverse modes of each scan.

### 1.3 Modelling the Spectra:

The VSTAR (Versatile Software for Transfer of Atmospheric Radiation) code is capable of predicting planetary spectra by combining line by line molecular and atomic absorption with full multiple scattering to solve radiative transfer (Bailey and Kedziora-Chudzer, 2012). Here we use VSTAR to model the transmittance of the radiation through the Earth's atmosphere and then to model the emission spectra of the Venus night side.

*Terrestrial transmittance spectra*
A standard 50 level mid-latitude clear sky terrestrial atmosphere was taken from the Intercomparison of Radiation Codes in Climate Models (ICRCCM) program. The tropospheric layers were then adjusted for local Siding Spring altitude and seasonal conditions and increased abundance of $CO_2$ to match current levels. The model includes molecular absorption from $CO_2$, $CO$, $CH_4$, $N_2O$, $H_2O$, $O_2$ and $O_3$ with absorption line information taken from HITRAN 2004 line database (Rothman *et al.* 2005).

*Venus night side spectra*

Our simulations of the Venus spectra are based on an atmospheric structure similar to that described by Meadows and Crisp, (1996). Temperature and pressure profiles for the lowest 100 km of the Venus atmosphere are obtained from the Venus International Reference Atmosphere (VIRA) (Seiff. 1985). Atmospheric gas molecules included in these studies are $CO_2$, $H_2O$, HDO, HF, HCl, and CO with the nominal mixing ratios obtained from previous ground-based observations by Bézard *et al.* (1990), De Bergh *et al.* (1995) and Pollack *et al.* (1993). Line databases were originally constructed for the terrestrial atmosphere and are recently becoming more relevant to the Venus troposphere as high temperature line positions and absorption strengths are being included to match the higher temperatures (737 K) and pressures (95.0 bar) of the Venus atmosphere (Donahue and Russell, 1997). Therefore the most complete line databases for each molecule are utilised in this study to maximise the accuracy of our models. $CO_2$ information was taken from the high temperature $CO_2$ line list described by Pollack et al. (1993). Based on the analysis and assessment of Bailey (2009), $H_2O$ lines were obtained from the BT2 database (Barber *et al.* 2006) and HDO from a similar database VTT (Voronin *et al.* 2010). HDO is significant in the Venusian lower atmosphere as it is enhanced $100 - 150$ times that of the terrestrial abundance. All other gases included are taken from the HITRAN 2004 database (Rothman *et al.* 2005).

There are four identified modal distributions of $H_2SO_4$ particles in the atmospheric haze and clouds in the Venus atmosphere, each with different mixing ratios for the relevant atmospheric layers. All four modes of $H_2SO_4$ particles are included in calculating the scattering, with information on the size, shape and vertical distribution and abundances obtained from Crisp (1986), based on information from both of the Pioneer Venus spacecraft (Ragent and Blamont 1980) but with the middle and lower cloud optical depths modified to fit ground-based AAT observations (Crisp *et al.* 1991). The surface is assumed to be Lambertian and to have an albedo of 0.15.

The radiative transfer solution at each wavelength was obtained by an eight stream discrete ordinate method using the DISORT software (Stamnes *et al.* 1988). The simulated Venus spectrum is adjusted for Doppler shifting before being multiplied by the terrestrial transmittance spectra. Only then are the spectra binned and convolved to our observed spectral resolution and bin locations. Simulations for Venus water vapour abundance profiles with increased water vapour in the troposphere and for different surface altitudes are run. The model output is then fitted to the corresponding observed spectrum to yield the water vapour abundance profile. The fitting is done using a simple least squares calculation between observed spectra and the scaled model spectra. The scaled model with the smallest residual is referred to as the best fit.

Modelling the 1.18 μm window is hampered by several factors which are explored in more detail below and which include: 1) an unknown source of $CO_2$ continuum absorption (Pollack *et al.* 1993), 2) lineshape variations due to weak overtone and hot $CO_2$ bands caused by the high temperatures and pressures near the Venusian surface and 3) line list completeness for $CO_2$ and $H_2O$ at the high Venus temperatures and pressures (Pollack *et al.* 1993, Bailey 2009). Another factor that could alter these results is 4) uncertainty in the water vapour continuum absorption. Recent measurements by Ptashnik et al. (2011) indicate that the continuum could be up to a factor of 10 larger than that assumed here.

An unknown continuum absorption that scales with the square of the $CO_2$ density is required to fit the observed spectra and is significant in $CO_2$ spectra at Venus surface pressures and temperatures (Pollack *et al.* 1993). This may be due to absorption from pressure induced bands of $CO_2$ and absorption from the very far wings of strong bands of $CO_2$ (Bézard *et al.* 1990). This additional absorption is compensated for in the model using 1 of 2 methods: 1) Lorentzian lineshapes adjusted by a $\chi$ factor out to short distances from the line centre (200 $cm^{-1}$) with the inclusion of an additional background continuum or 2) Lorentzian lineshapes that are adjusted by a $\chi$ factor over large distances (1000 $cm^{-1}$) such that it tends towards a constant rather than zero and therefore inclusively compensates for the continuum absorption, based on work by Perrin and Hartmann (1989). The lineshape used for the *VSTAR* modelling of $CO_2$ in the Venus atmosphere is that described by Meadows and Crisp (1996) and is a modification of that by Perrin and Hartmann (1989).

Emission angle variations (as reported in Bézard *et al.* 2009), surface emissivity, variations in cloud particle size and lapse rate variations produce a flux variation in our modelling without changing the spectral shape of the window. As we determine the Venusian water vapour abundance by matching spectral ratios and window gradients, we can ignore the above parameters in this study.

Line databases were initially generated to model the Earth's atmosphere and omit hot bands due to high energy transitions that are important in hotter atmospheres such as Venus. Errors in $CO_2$ and $H_2O$ line lists affect the inferred water vapour abundance. Bailey (2009) models the observed spectra presented in Meadows and Crisp (1996) using the more accurate and more complete BT2 line lists and found that the Meadows and Crisp value of 45 +/- 10 ppmv for water vapour mixing ratio reduced to 27 +/- 6 ppm. Bailey (2009) compared several line lists and determined that the BT2 (Barber *et al.* 2006) line list is the most complete and accurate representation of the $H_2O$ spectrum at Venus temperatures. Similar issues exist for $CO_2$ line lists contained in HITRAN and GEISA. A larger more comprehensive list the Carbon Dioxide Spectroscopic Databank (CDSD; Tashkun *et al.* 2003) also exists, however, it is known to have missing transitions over our spectral region of interest (CDSD, 2010; Bézard *et al.* 2011) and therefore was not used for these studies. We therefore use the high temperature $CO_2$ list utilised by Pollack *et al.* 1993. Current line lists include air broadening and self broadening line half widths, but $CO_2$ broadening in an atmosphere like Venus will result in somewhat different line widths for minor species. Data on $CO_2$ broadening is not available for all species. Delaye *et al.* (1989) calculated $H_2O$ line widths broadened in $CO_2$ and temperature coefficients for a range of values of rotational quantum numbers, albeit the transition lists were not complete. Delaye's data has been shown to agree well with measured broadening parameters (Langlois *et al.* 1994). $CO_2$ broadened $H_2O$ lines were calculated in *VSTAR* using information from Delaye *et al.* (1989). For all other minor gases air broadened values obtained from the respective databases were used.

We chose four spectral regions in the 1.18μm window for our spectral fitting shown in figure 3. Three over individual features (f1, f2, f3) and a fourth broad peak (pk) that covered several water vapour features and included the continuum gradient shown in figure 3.

A sensitivity profile for the Venus model was generated by varying the water vapour abundance for each individual 2 km altitude interval by 20 %, whilst keeping all the other altitude intervals at the nominal level. This was repeated until all the levels have been varied. RMS errors were then found for each output spectrum with respect to the nominal output spectrum for each of the four match regions (figure 4). Figure 4 indicates that the Venus model has a maximum sensitivity across the peak match region at 12 km altitude (4–27 km half power range) and slightly higher maximum sensitivities for the features 1, 2 (15 km) and 3 (14 km) with half power ranges of 6–31 km, 7–35 km and 5–40 km respectively. This also indicates that utilising individual spectral features (such as f1) can result in higher sensitivity studies of water vapour in the lower Venus atmosphere.

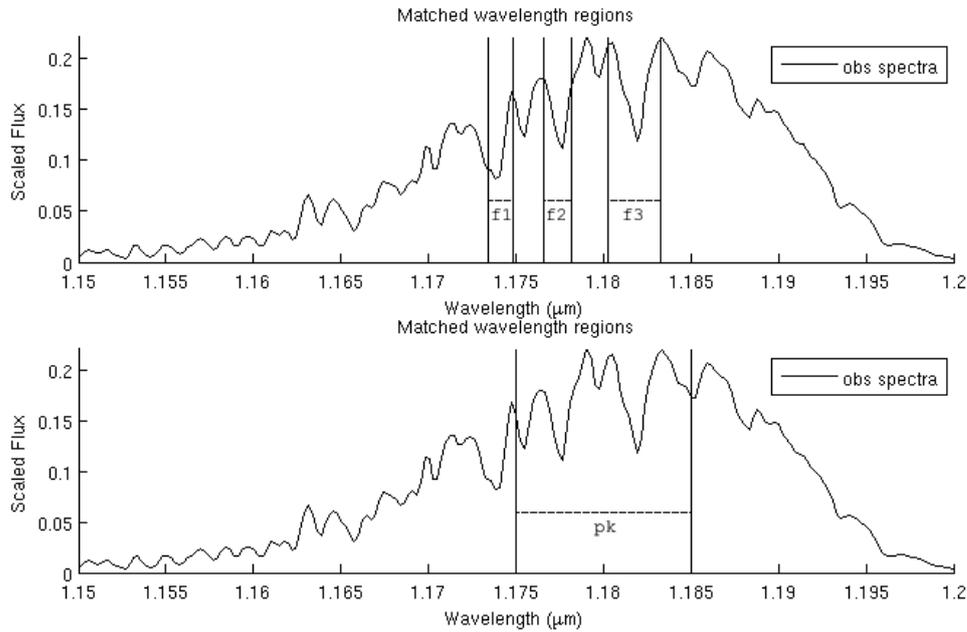

Fig 3: Observed spectra of the Venus night side with lines indicating the spectral match ranges chosen for this study: f1: near 1.174 μm, f2: near 1.177 μm, f3: near 1.182 μm and pk: including several water vapour absorption features across the central peak of the 1.18 μm window.

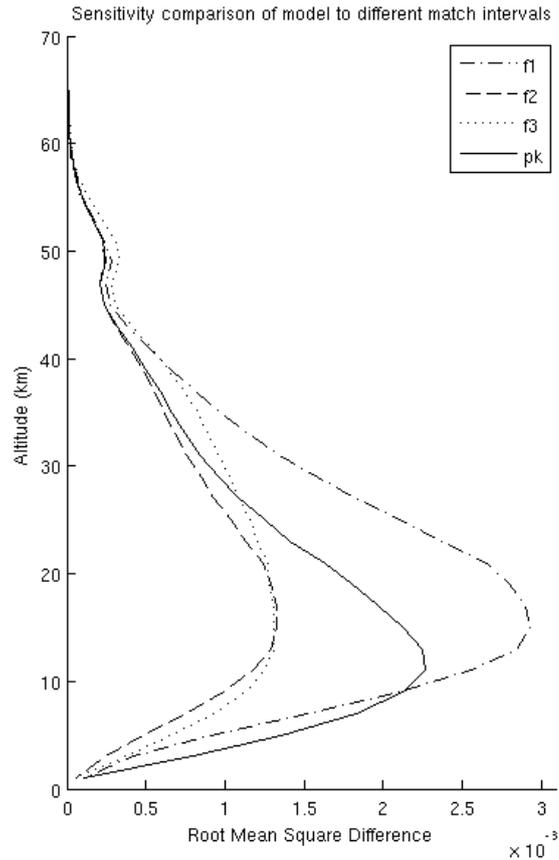

Fig 4: Shows the sensitivity of water vapour variations with altitude for each of the match regions used in this study: pk, f1, f2 and f3.

The modelled water vapour profiles between the altitudes of 100 km and 55 km are the same for all profiles being considered and fitted (see fig 5). The water vapour abundance increases with decreasing altitude at a similar rate for all water vapour profiles until the maximum abundance for each model is reached and that model then remains constant to the surface. Measurements of the 2.3 µm window indicate that an average water vapour abundance of 30 ppmv at altitudes of 30 – 40 km (Marcq, et al 2008), therefore all water vapour profiles that are generated to reach abundances greater than 30 ppmv are forced to reach 30 ppmv by 40 km altitude for consistency. All profiles have reached their maximum by 28 km and all profiles are therefore constant across the altitudes of peak sensitivity.

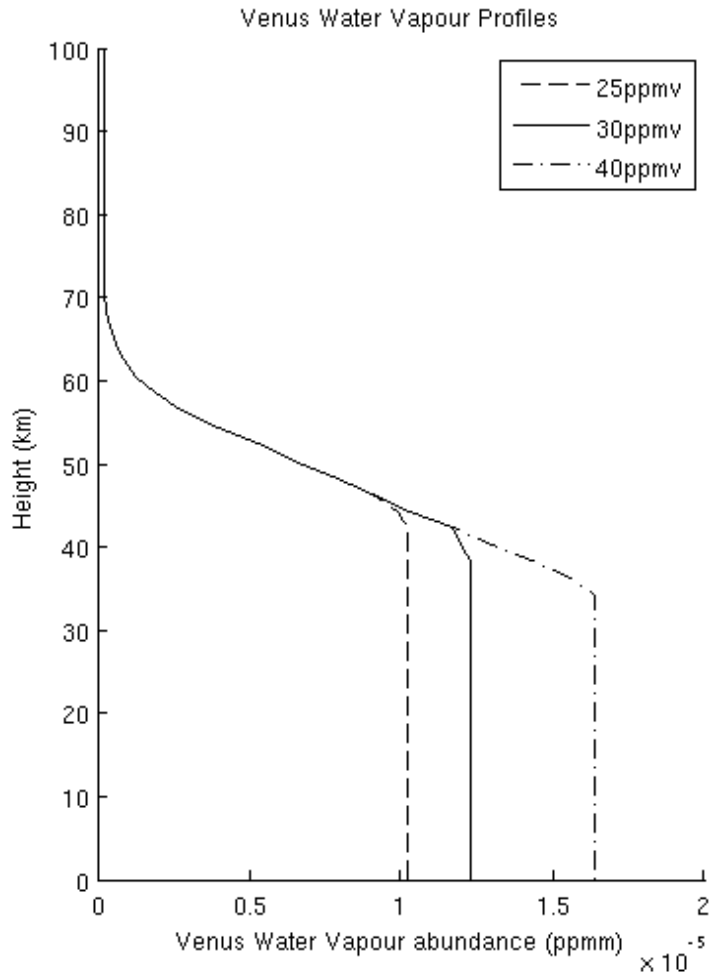

Fig 5: The Venus water vapour profiles associated with 25ppmv, 30ppmv and 40ppmv. The x axis is given in parts per million mass.

## 1.4 Modelled Results

Observed spectra are extracted in a grid pattern across the Venus night side disk. Spectra obtained too close to the crescent are noisy due to remanent effects of the scattered light and are removed from the study. Weak spectra too close to the limb of the planet are radially excluded (fig 6).

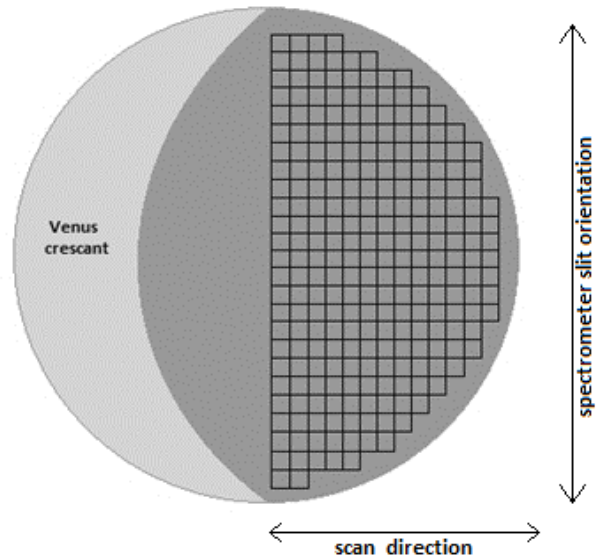

Fig 6: Shows the Venus disk and the spectral extraction location with respect to the edge of the disk and the Venus crescent. Each square in the grid is equivalent to a 3 by 3 pixel region in the database used for this study.

The surface elevations corresponding to the region where each of the extracted spectra were obtained are determined by matching the observed pixel location to the Magellan topographic map and taking an average of altitudes over the projected pixel area. The extracted spectrum is fit to modelled spectra by varying the Venus water vapour abundances until a best fitted is found. The spectral wavelength range of the match regions are given in Table 3.

Table 3: shows the wavelength range for each spectral match regions.

|  | Match range start wavelength | Match range end wavelength |
|---|---|---|
| Peak | 1.1750 µm | 1.1850 µm |
| Feature 1 (f1) | 1.1734 µm | 1.1748 µm |
| Feature 2 (f2) | 1.1766 µm | 1.1782 µm |
| Feature 3 (f3) | 1.1803 µm | 1.1833 µm |

Various Venus water vapour abundances are modelled in order to observe the differences across the match regions selected (fig 7). All of the individual water vapour features f1, f2, and f3 are seen to deepen as the Venus water vapour increases. The peak (pk) match region also shows that in addition to the above, the relative gradient across the 1.18 µm window becomes steeper.

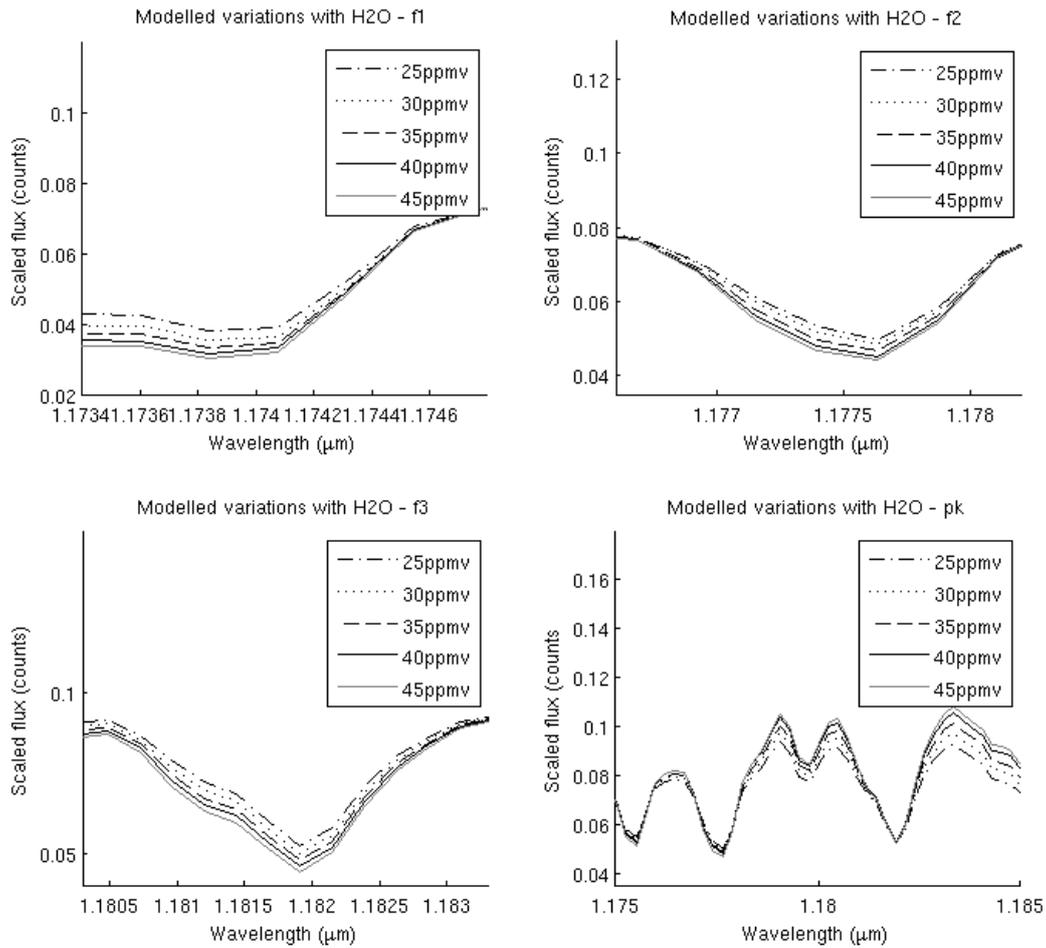

Fig 7: Shows the modelled variations in spectra caused by increasing Venusian water vapour abundances for each of the spectral match regions pk, f1, f2, and f3.

Various Venus water vapour abundances are modelled in order to observe the differences across the match regions selected (fig 7). All of the individual water vapour features f1, f2, and f3 are seen to deepen as the Venus water vapour increases. The peak (pk) match region also shows that in addition to the above, the relative gradient across the 1.18 μm window becomes steeper.

Figures 8 -11 show best fit water vapour abundance for each extracted spectra from both scans 222 and 523. The best fit water vapour is found to be 31 ppmv taken from the pk match region across the centre of the 1.18 μm window and feature f3, whereas the spectral features f1 and f2 have higher best fit values 34 ppmv and 37 ppmv. The best fit abundance scatter is larger for the fits to water vapour features than that for the peak fit. The range and standard deviation for each match region are shown in Table 4.

Table 4: Best fit abundances for each match region pk, f1, f2 and f3 and their respective best fit range and the standard deviation.

| Match Regions | Best Fit water vapour abundance (ppmv) | Scatter Range (ppmv) | Standard Deviation (ppmv) |
|---|---|---|---|
| Pk | 31 | 25 − 40 | 2.15 |
| F1 | 34 | 21 − 50 | 4.50 |
| F2 | 37 | 17 − 58 | 6.88 |
| F3 | 31 | 18 − 49 | 4.22 |

The standard deviations are compiled from the scatter in each diagram (figs. 8 -11). Dashed lines on the top left plot of each figure show the best fit value of 30 ppmv and associated error bars (+10 − 5 ppmv) from the recent analyses of SPICAV/VEX data (Bézard *et al.* 2011) for comparison. The scatter graphs indicate that fits to the pk match range are much tighter than those obtained across individual absorption features. Scatters also appear to be broader across f2 compared to f1 and f3.

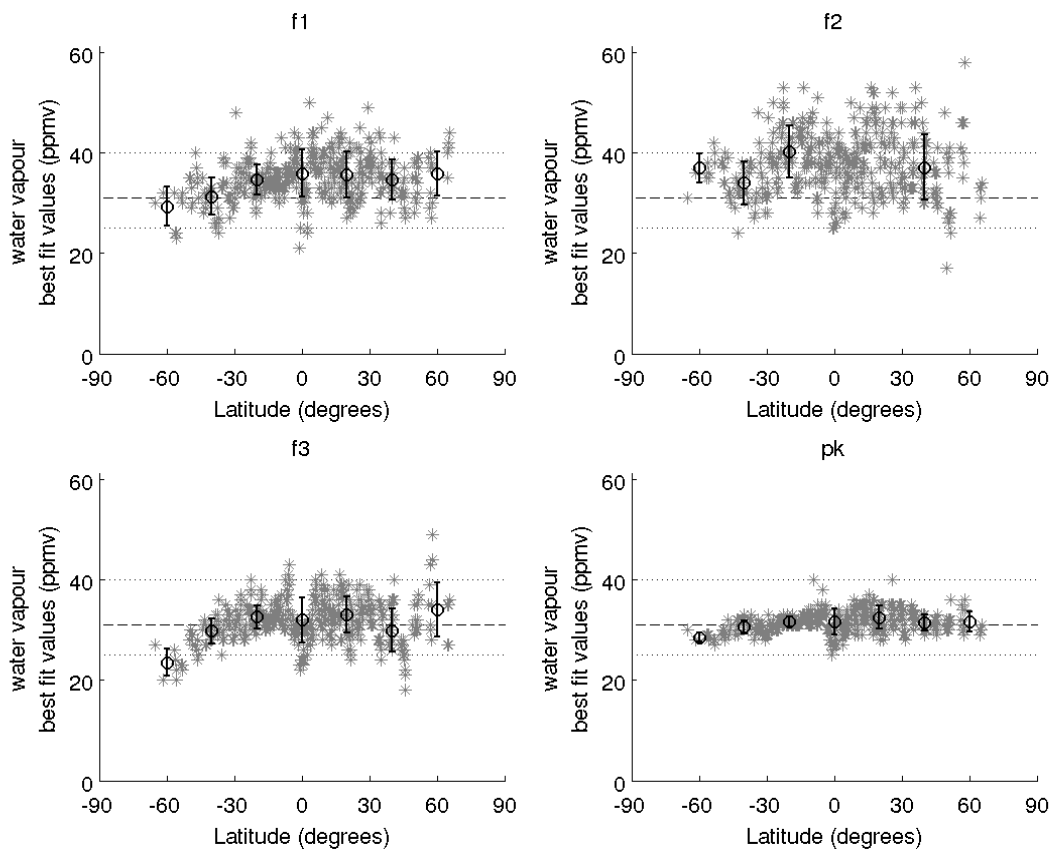

Fig 8: Best fit water vapour abundances (stars) plotted against latitude (degrees). Binned averages and standard deviations are shown (black). For comparison, the water vapour average corresponding to the pk wavelength region is provided on all plots (dashed line) and associated spread (dotted lines).

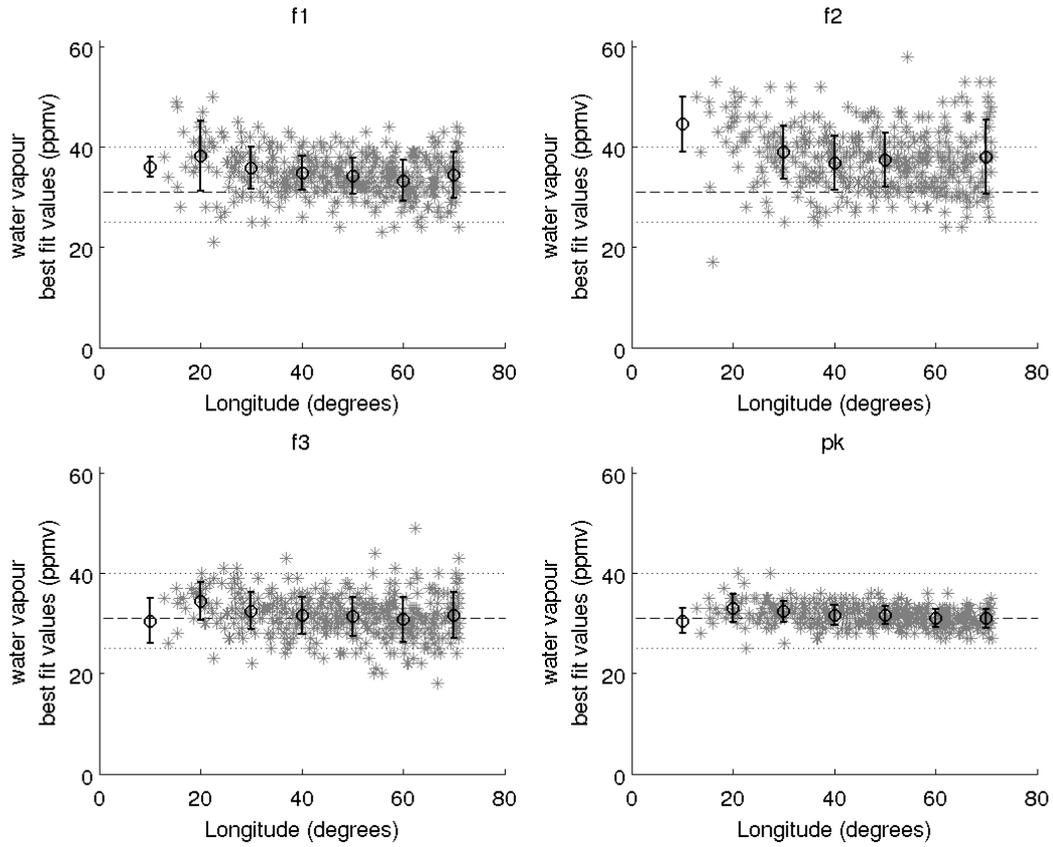

Fig 9: Best fit water vapour abundances (stars) plotted against longitude (degrees). Binned averages and standard deviations are shown (black). For comparison, the water vapour average corresponding to the pk wavelength region is provided on all plots (dashed line) and associated spread (dotted lines).

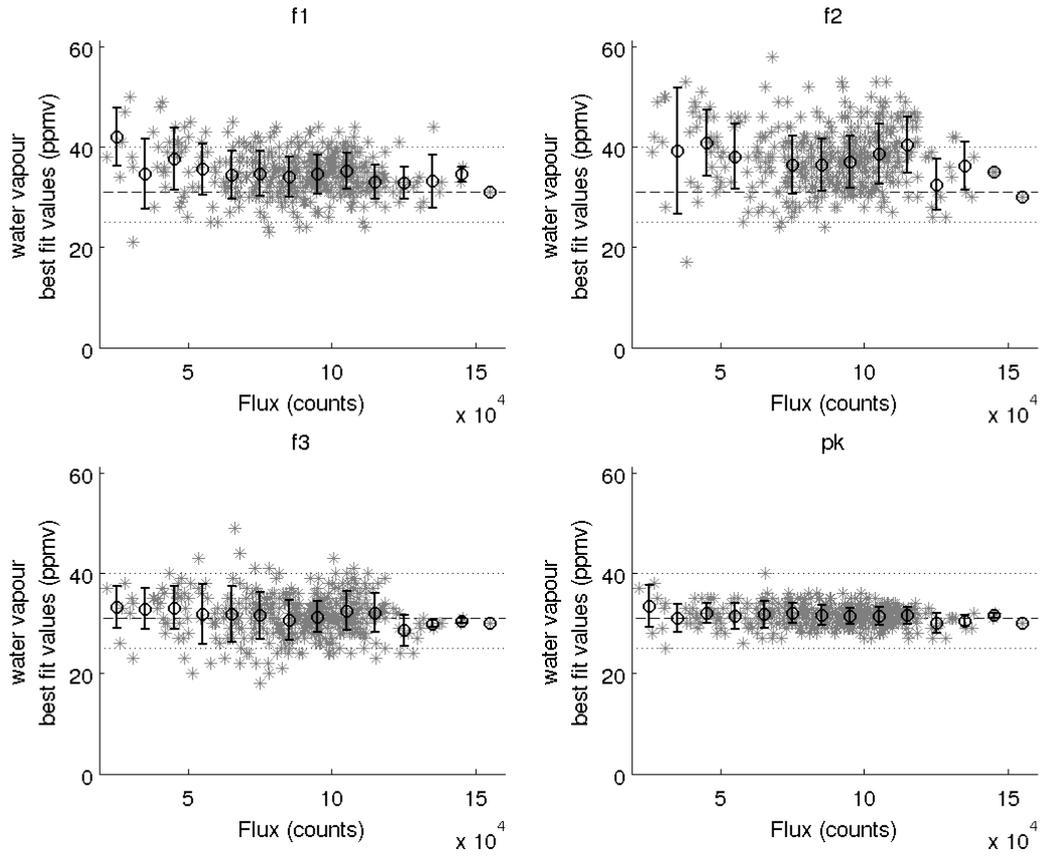

Fig 10: Best fit water vapour abundances (stars) plotted against flux (counts). Binned averages and standard deviations are shown (black).

For comparison, the water vapour average corresponding to the pk wavelength region is provided on all plots (dashed line) and associated spread (dotted lines).

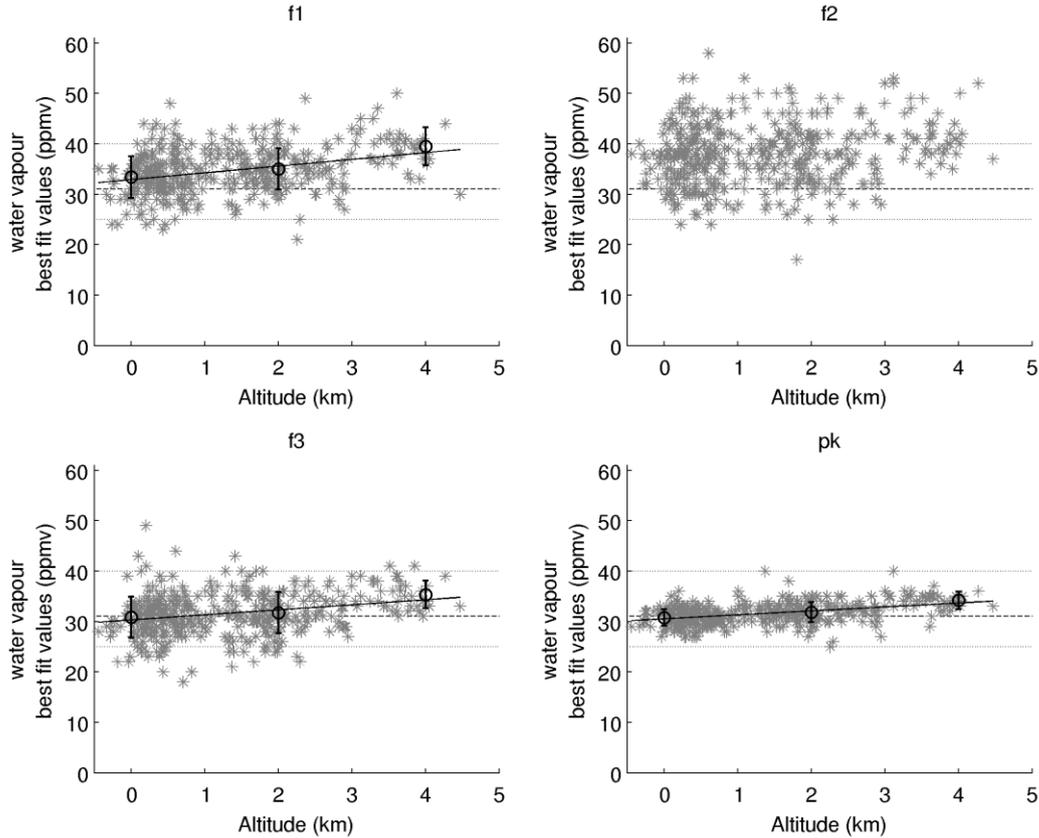

Fig 11: Best fit water vapour abundances (stars) plotted against surface altitude (km). Binned averages and standard deviations are shown and a best fit 1 degree polynomial is fitted to the data (black). For comparison, the water vapour average corresponding to the pk wavelength region is provided on all plots (dashed line) and associated spread (dotted lines).

Figures 8 and 9 show variations with best fit abundance plotted against a spatial axis (latitude and longitude respectively). Figure 8, in the plots for f1, f2 and pk, decrease in water vapour abundance from the equator towards the southern pole. No corresponding decrease is observed towards the northern pole nor with longitude (Fig. 9) indicating that this is unlikely to be viewing angle error. This may be a reflection of true water vapour variations with latitude however the decrease is only minor for the f1 and pk plot which are the most sensitive to water vapour variations (Fig. 4). Figure 9 shows no observed spatial variations in longitude within the limits of our data. Figure 10 shows the best fit water vapour abundance against observed flux, and therefore cloud opacity which is the major source of opaqueness in the Venusian night side emission. No trend with flux across the night side disk is observed in our dataset. Figure 11 shows variations with near surface water vapour abundance (as outlined by surface topography). Each plot shows a clear trend of increasing water vapour abundance with surface elevation. These results are discussed further in section 1.5.

## 1.5 Discussion
### 1.5.1 Best fit abundances

Best fit abundances and error bars obtained from fits to the pk match region agree closely with recently published results by Bézard *et al.* (2011) of 30 ppmv (+10/-5 ppm) using the higher spectral resolution instrument SPICAV/VEX and disagree with the earlier results published by Bézard et al. (2009) of 44 ppmv (+/-9 ppm) using the lower spectral resolution instrument VIRTIS-M/VEX. The best fit abundances obtained from individual spectral features f1, f2 and f3 have higher fit values and/or higher scatter. A comparison within each scatter graph of figures 8 -11 shows that f2 returns on average the highest fit value and scatter.

The wider scatter associated with individual water vapour feature matches may be due to several reasons: lack of additional window continuum constraints, noise and inadequately resolved features. The water vapour abundance not only affects the depth of the absorption features but also the gradient of the window, as shown in Fig 7. In the evaluation of the pk region the model is fit primarily to the window gradient and secondly to the absorption features. Therefore the resulting fit may not fully compensate for the whole contrast of the absorption features. There are fewer spectel bins across the match regions f1, f2 and f3 and without the continuum constraint the model is free to increase or decrease the water vapour abundance, producing a larger scatter. Noise will therefore have a greater influence on spectral fit for f1, f2 and f3. We note that feature 2 has the largest scatter in figs 8 -11. Figure 4 indicates that this match region is least sensitive to water vapour variations at most altitudes, and therefore will have a higher scatter range. Fig 7 shows that the largest water vapour induced spectral gradient occurs across feature 2. If this spectral gradient is not taken into account during the evaluation of water vapour abundance, a larger scatter may be induced. This issue is also shown by the mismatch between observed and modelled spectra near 1.179 μm in Fig. 12 (f2 plot) and 1.183 μm in Fig. 12 (f3 plot). A spectral mismatch caused by missing absorptions in the $CO_2$ and/or $H_2O$ line lists can also induce a scatter in the determined water vapour abundance if missing or incorrect absorptions occur at corresponding wavelengths to a water vapour absorption feature. Fig 12 (bottom) shows the variation in best fit model spectrum matches when fitted to the same observed spectra across different water vapour features.

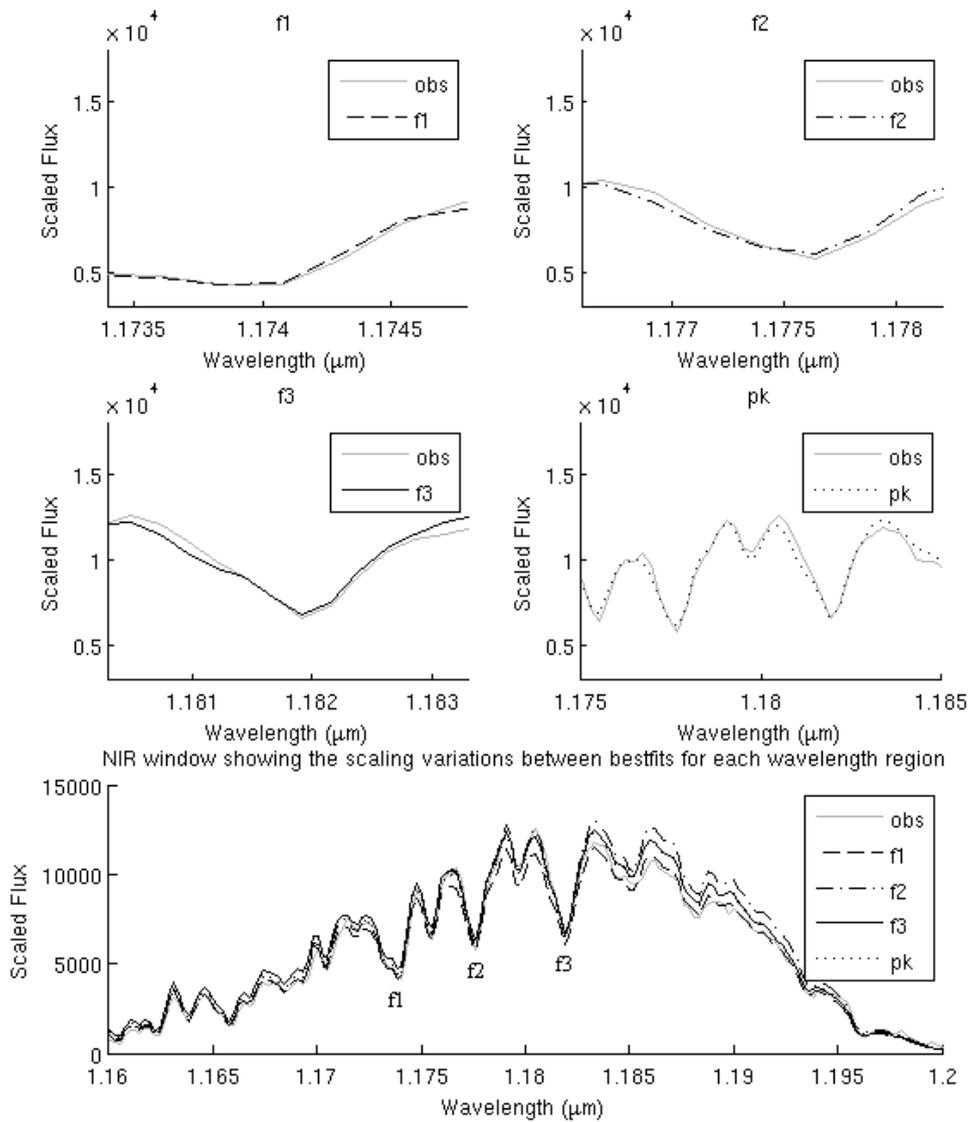

Fig. 12: Venus observed spectra are fitted to a scaled modelled spectrum. The top four plots show the best fit scaled model for each wavelength match region. The bottom plot shows how each best fit scaled model fits the whole 1.18 μm window.

Fig 13 displays a comparison between the spectral resolution of this study (R~2400) and a higher resolution spectrum (R>100000). Variations in water vapour abundance, temperature or pressure have different effects on the shape as well as the depth of each absorption feature. These effects are not fully observed in our data, contributing to the higher scatter between best fit abundances for absorption feature fits f1, f2 and f3.

Mismatches between the modelled and observed spectra are likely due to one or several of the following problems; line list completeness, absorption line shape and background continuum absorption; none of which can be independently determined from water vapour abundance using our current data. Fully resolved spectral observations across this window region may be needed to resolve these modelling issues.

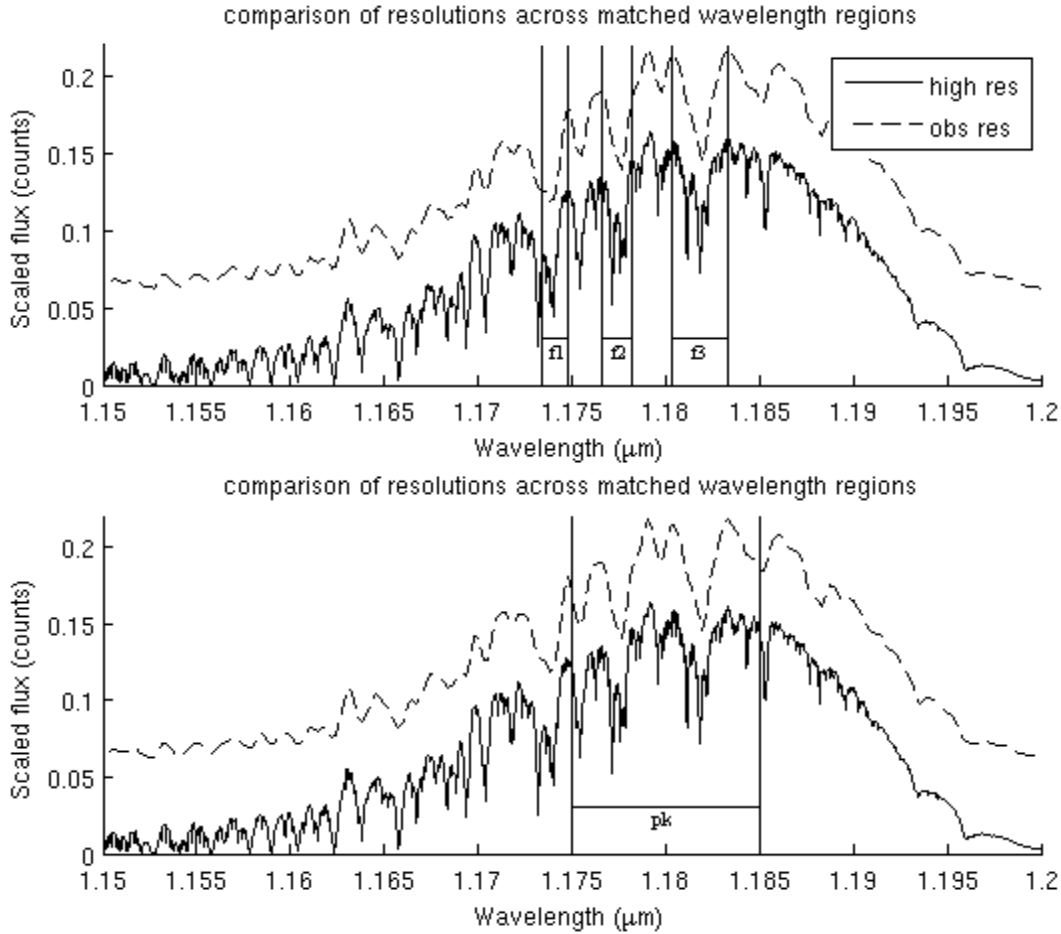

Fig 13: The modelled 1.18µm spectra at the spectral resolution of this study (dashed line) compared to a model of the same region at a higher spectral resolution output (solid line). The spectra have been offset vertically from each other for ease of comparison.

### 1.5.2 Spatial variations

Spatial variations were found by Tsang et al. (2010) from studies of water vapour in the 2.3µm window, which probes higher altitudes (30 - 40km). These variations were found to be correlated to cloud opacity. Barstow et al. (2012) found similar correlations at 40km) under strong cloud opacities but found the correlations to be weaker for regions of low cloud opacities. Our results indicate that there are no correlations between water vapour abundance and flux (and therefore cloud opacity) observed within the limits of our data and at our spatial resolution (Fig. 10) for altitudes 12 – 20 km.

Prior observations of the 1.18 µm window have found no evidence of spatial variations in the detected water vapour abundance. Bézard et al. (2009) placed an upper limit on variations in $H_2O$ abundance using the 1.18 µm window (15 – 20km altitude) of +/-1% between 60 S and 25N and +/-2% for 80S to 25N derived from VEX/VIRTIS-M data. Figure 8 shows a small decrease in water vapour abundance between the equator and the south pole of a few percent. The lack of corresponding variations in longitude, or mirrored decrease between the equator and north pole would seem to rule out errors due to viewing angles and surface elevation effects (as described in section 1.5.3).

### 1.5.3 Near surface water vapour abundance

Figure 11 shows a trend of increasing water vapour abundance with surface altitude. We note that Figure 4 shows that the 1.18 μm window is only weakly sensitive to altitudes below 5 km, however the trend is strongest for feature 1 and peak match ranges which are most sensitive to variations in water vapour closest to the Venus surface. The correlation coefficients for features 1,2,3 and the peak range are all weak (between 0.13 and 0.43), but the p-values for all plots within fig. 11 are statistically significant, within the significance value $\alpha = 0.01$. Whilst this means that a null hypothesis is unlikely it does not rule out that the correlation may be caused by the elevation dependence of surface temperature, variations in composition or systematic errors in the poorly-constrained radiative transfer properties of the lowest atmospheric layer. Altitude binning errors are one possibility that may also be influencing this gradient. Modelled altitudes are sampled every 2 kms (0 km, 2 km and 4 km) such that each observed pixel spectrum is matched to a model spectrum generated from the closest altitude bin. As lower altitudes have longer path lengths, a spectrum from slightly higher altitudes to that of the bin average altitude will plot at a lower water vapour abundance and a spectrum from slightly lower altitudes to the bin average altitude will plot at a higher water vapour abundance. There will be fewer pixels binned from below to 0 km than from above and vica-versa for 4 km. These errors would produce an apparent decrease in average water vapour for 0 km altitudes and an apparent increase in average water vapour for 4 km altitudes, mimicking the gradient shown in figure 11. However, this effect would mean that the points that fall near the binned average altitude would plot the correct water vapour abundance and a best fit line drawn between the 3 points should fall at the same abundance if no real water vapour gradient is present. Also this effect should cause a corresponding decrease in water vapour with altitude across each 2 km bin, which we do not see in figure 11. Future higher spectral and spatially resolved data analyses should be made with a greater number of altitude bins to reduce this effect.

Trends in the near surface water vapour absorption are further studied by utilising spectral ratios. A spectrum obtained over a surface altitude of 4 km is divided by a spectrum obtained over a surface altitude of 0 km. The resulting spectral ratio shows the $CO_2$ and $H_2O$ absorptions solely due to the atmosphere between these two altitudes. Fig 14 shows a topographic map of Venus with extraction locations for Fig 15 (Fig 14a) and Fig 16 (Fig 14b).

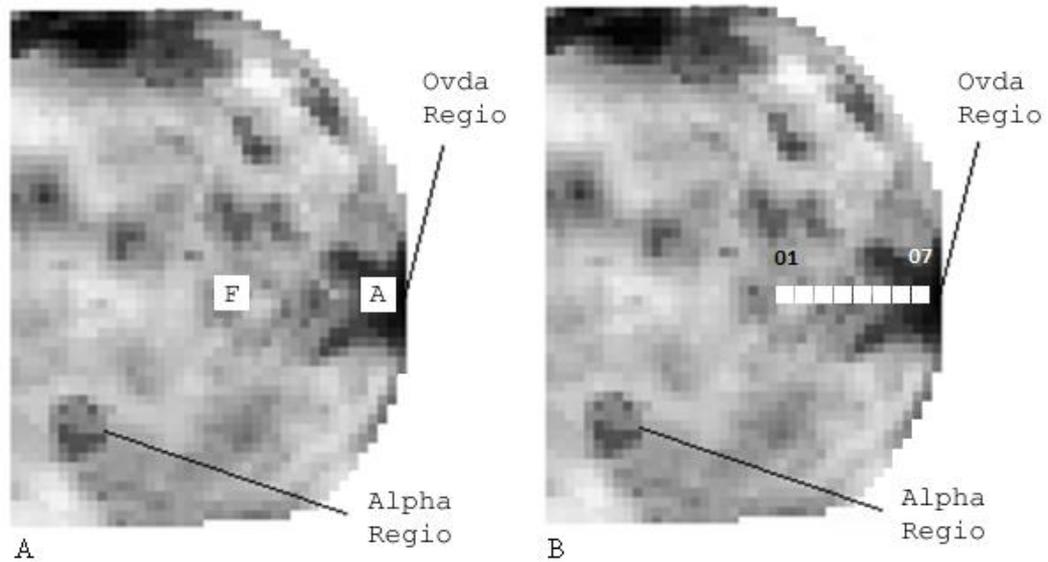

Fig 14: A binned Magellan map of Venus topography oriented to match the topography of Venus at the time of our observations. Panel A shows two 6 by 6 pixel squares that were used to produce Fig 15. The square labelled A refers to a region across Ovda regio with an average surface altitude of about 4 km. The square labelled F refers to a region with an average altitude of approximately 0 km altitude. Panel B shows a series of 3 by 3 pixel squares used to produce Fig 16. The square labelled 01 has an altitude of approximately 0 km, each square subsequent is over a region of increasing surface altitude reaching a maximum at the square labelled 07 which has an average altitude of 4 km.

Two 6 by 6 pixel squares were obtained from over Ovda Regio (average 4 km elevation) and from a reference region (average 0 km elevation) (see fig 14a). The spectral ratio of these extracted spectra is compared to modelled spectral ratios in Fig 15. Vertical lines indicating the central position of spectral features f1, f2 and f3 are provided (fig. 15) and show the water vapour absorption features due only to the lowest 4km in the Venus atmosphere. Figure 15 indicates that 31 ppmv is consistent with measurements to the surface, however the signal to noise ratio for this spectrum is not high enough to detect the existence of a near surface water vapour gradient. Should this gradient be real however, this would imply a chemical reaction between the surface rock and the water vapour in the atmosphere.

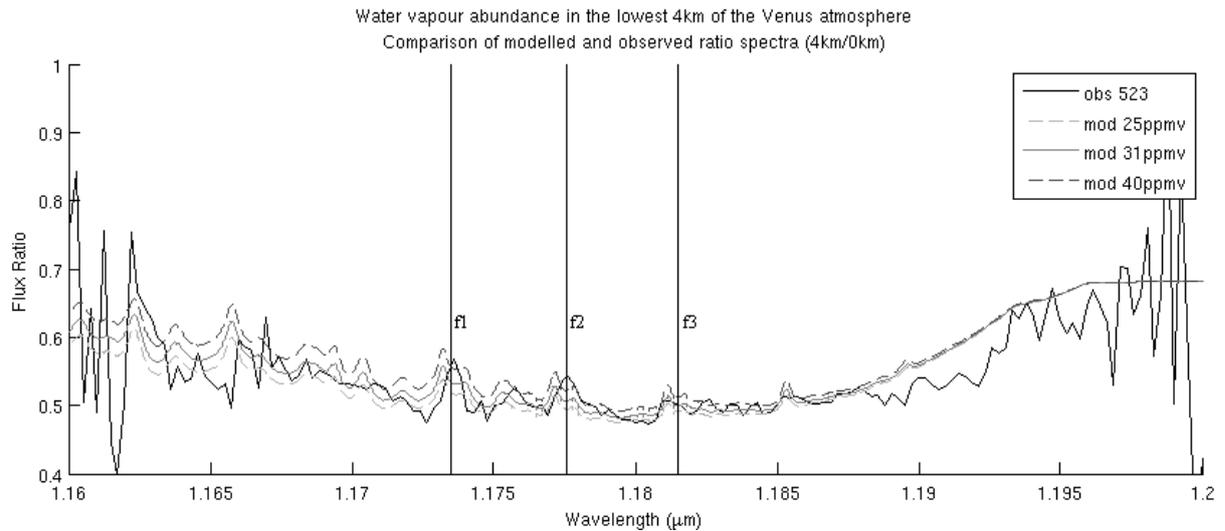

Fig 15: The observed spectra presented here (black line) is the ratio of a spectrum obtained from 4 km elevation (Ovda Regio) divided through by a spectrum obtained over surface altitude of 0 km elevation. The modelled spectra (grey lines and dashed lines) are simulated over the same altitudes and through corresponding air masses to the observed spectral ratio. The modelled spectra depict different water vapour profiles (25, 31 and 40 ppmv).

Figure 16 demonstrates a potential method to outline the $H_2O$ profile of the Venus near surface troposphere. A series of adjacent 3 by 3 pixel squares obtained over the Ovda Regio rise are divided through by a common spectrum (fig. 14b). Variations between the spatial gradient of these spectral ratios show the relative signatures of water vapour and $CO_2$ between the different altitudes. The spectral absorption features are not identified above the noise in these spectra in Fig 16, however the gradient across the 1.18 µm window is seen to flatten as the difference between the altitudes decreases between the ratio spectra. This shows the changing influence of $CO_2$ and $H_2O$ absorptions due only to the varying altitude regions between the spectra. Further work is required to model and interpret the abundance profile using this method.

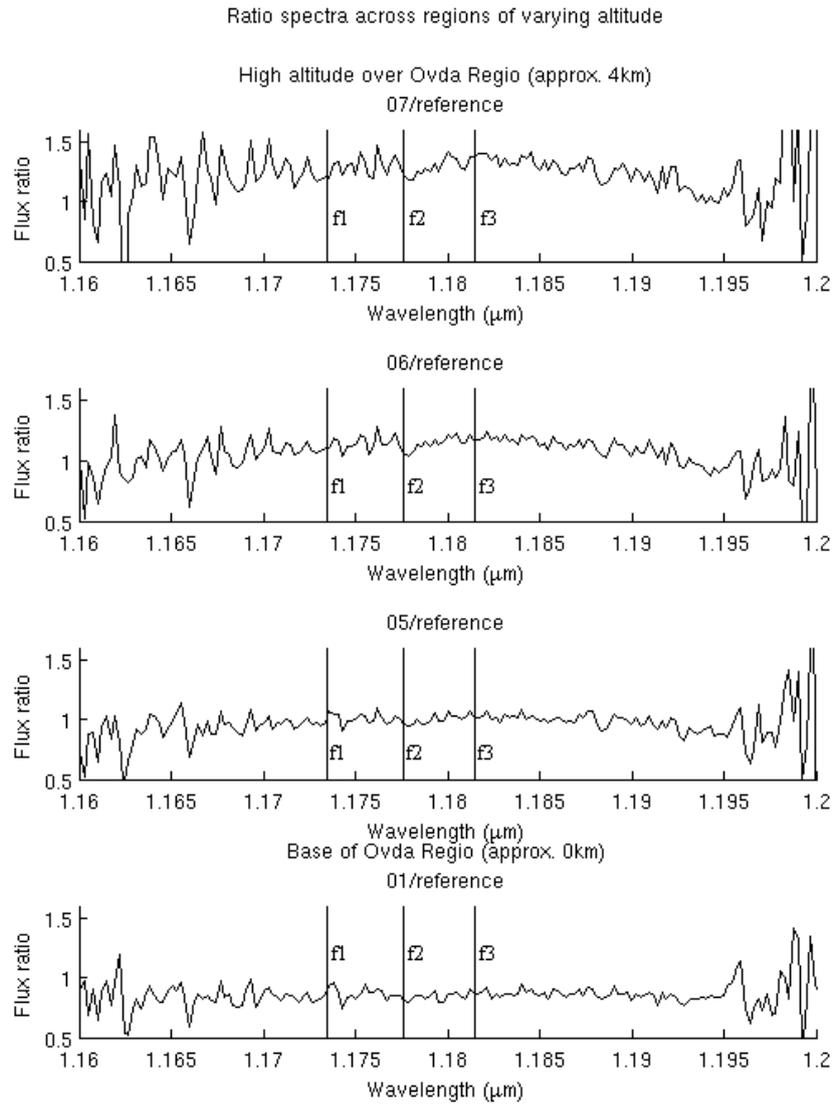

Fig 16: A series of flux ratios are shown. 07 is obtained from the highest point of the Ovda Regio rise (4 km above the reference spectrum) and 01 at the lowest end of the rise (at a similar altitude to the reference spectrum) (see fig 14b for numerical references). The spectral positions of water vapour absorptions are indicated by the vertical lines. The gradient variations in the spectral ratios at the short wavelength side of the window corresponds primarily to the change in water vapour abundance between different altitudes, whereas variations observed in the long wavelength side of the window correspond mainly to variations in $CO_2$ between the different altitudes. The flattening of the gradients across the window with decreasing separation altitude between the spectra and the reference spectra show the influence of $CO_2$ and $H_2O$ between the corresponding altitudes.

### 1.5.4 Future directions

Further study is required to determine the near surface water vapour profile. It may be possible to determine the water vapour gradient by spectrally fitting different wavelength ranges within the 1.18 μm window. Meadow and Crisp, (1996) and Pollack *et al.* (1993) show that the peak of the 1.18 μm window is sensitive to slightly lower altitudes than those of the window wings, and this is also confirmed in Fig. 4. Studies at much higher spectral resolutions could fully resolve individual absorption lines which provide molecular rotational information dependent on temperature and pressure within the Venus atmosphere. This would allow for better constrained identification of the water vapour profile at altitudes between 12 km and 20 km. This has not been achieved with our current dataset due to the limited spectral resolution. Studies with much greater signal to noise data can then also be used to determine the water vapour abundance over surface topography. Radiative transfer forward modelling has been a successful tool for the interpretation of Venus night side spectra and higher spectral resolution observations will also help to further constrain errors associated with the $CO_2$ lineshape and the completeness of current $CO_2$ and $H_2O$ line list databases.

## 1.5 Conclusion

Ground-based near-infrared observations of water vapour in the Venus troposphere were obtained using IRIS2 on the Anglo-Australian Telescope at Siding Spring, Australia shortly after the June 2004 inferior conjunction. We have retrieved the mixing ratio and distributions of the water vapour abundance in the Venus lower troposphere as determined by modelling different wavelength regions within the 1.18µm window. We find a best fit water vapour abundance matched to the peak of the emission window of 31 ppmv - 6 + 9 ppmv which agrees with previously published results Bézard et al. (2011) using SPICAV/VEX in contrast to results produced using VIRTIS-M/VEX by Bézard et al. (2009) of 44 ppmv +/-9 ppmv. Fitting a broad wavelength peak (1.175 – 1.185 µm) is shown to be the method that provides the least scatter of fits at our spectral resolution. However we show that utilising individual water vapour features within the 1.18 µm can provide a higher sensitivity to water vapour abundance in the Venus lower atmosphere than using the peak emission (1.175 and 1.185 µm) and that each water vapour feature has a different altitude sensitivity that can be utilised in future studies with higher spatial and spectral resolutions. No variations in water vapour near 15 km altitude can be associated with cloud opacity and from the lack of $H_2O$ spatial variability, the troposphere at this altitude appears to be globally well mixed. A possible weak spatial gradient in water vapour on the order of a few percent is detected between the equator and the south pole and in the near surface water vapour profile, however no other gradients in water vapour are observed above the errors of our data. We have shown that we can detect near surface (0 km – 4 km) water vapour absorption through spectral ratios utilising variations in the surface altitude and have shown that a constant abundance of 31 ppmv is consistent with our data. However currently the signal to noise of our data is too low to neither negate nor confirm the existence of a possible water vapour gradient in the Venus troposphere. Future observations made with very high spectral resolution and sufficient signal to noise are required to verify the detections of the weak horizontal and vertical gradients.

## Acknowledgements
Many thanks go to Dr. David Luz for his review of this paper and helpful conversations and suggestions. We acknowledge FCT funding through project grants POCI/CTE-AST/110702/2009, Pessoa-PHC programme and  project PEst-OE/FIS/UI2751/2011.